\documentclass[aps,prd,twoside,twocolumn,nofootinbib,10pt,showpacs,floatfix]{revtex4-1}
\usepackage{amsmath,amssymb}
\usepackage{graphicx,bm}
\usepackage{slashed}
\usepackage{epstopdf}
\usepackage{ulem} 
\usepackage[usenames]{color}
\usepackage{float}
\usepackage{hyperref}
\usepackage{subfigure}
\usepackage{subfigure}
\usepackage{rotating}
\usepackage{color}
\usepackage{multirow}
\usepackage{dcolumn}
\usepackage{overpic}
\usepackage{booktabs}
\usepackage{makecell}
\usepackage{diagbox}

\renewcommand\sout{\bgroup \color{red} \ULdepth=-.5ex \ULset}

\newsavebox{\tablebox}
\begin{document}
\title{Emergence of molecular-type characteristic spectrum of hidden-charm pentaquark with strangeness embodied in the $P_{\psi s}^\Lambda(4338)$ and $P_{cs}(4459)$}
\author{Fu-Lai Wang$^{1,2,3}$}\email{wangfl2016@lzu.edu.cn}
\author{Xiang Liu$^{1,2,3}$}\email{xiangliu@lzu.edu.cn}
\affiliation{
$^1$School of Physical Science and Technology, Lanzhou University, Lanzhou 730000, China\\
$^2$Research Center for Hadron and CSR Physics, Lanzhou University and Institute of Modern Physics of CAS, Lanzhou 730000, China\\
$^3$Lanzhou Center for Theoretical Physics, Key Laboratory of Theoretical Physics of Gansu Province, and Frontiers Science Center for Rare Isotopes, Lanzhou University, Lanzhou 730000, China}

\begin{abstract}
Inspired by the newly observed $P_{\psi s}^\Lambda(4338)$ and the reported $P_{cs}(4459)$, we indicate the existence of the molecular-type characteristic mass spectrum for hidden-charm pentaquark with strangeness. It shows that the $P_{cs}(4459)$ may contain two substructures corresponding to the
$\Xi_c\bar{D}^*$ molecular states with $J^P=1/2^-$ and $3/2^-$, while there exists the corresponding  $\Xi_c \bar{D}$ molecular state with $J^P=1/2^-$. As the prediction, we present another characteristic mass spectrum of the $\Xi_c^\prime\bar{D}^{(*)}$ molecular states. Experimental confirmation of these characteristic mass spectra is a crucial step of constructing hadronic molecular family.

\end{abstract}
\maketitle

\section{Introduction}\label{sec1}

Since the birth of the quark model \cite{GellMann:1964nj,Zweig:1981pd}, the concept of exotic hadronic matter has been proposed, which has attracted extensive attention from both experimentalists and theorists. Especially, with the accumulation of experimental data, the observations of charmoniumlike $XYZ$ states and $P_c$ states in the past around twenty years make this issue becomes hot spot of hadron physics up till now \cite{Liu:2013waa,Hosaka:2016pey,Chen:2016qju,Richard:2016eis,Lebed:2016hpi,Brambilla:2019esw,Liu:2019zoy,Chen:2022asf,Olsen:2017bmm,Guo:2017jvc,Meng:2022ozq}. These studies are helpful to deepen our understanding of non-perturbative quantum chromodynamics (QCD). Among different exotic hadronic matters, the hadronic molecular state was popularly applied to explain these novel phenomena relevant to new hadronic states since most of observed hadronic states are close the thresholds of hadron pair. A typical example is the LHCb's discoveries of the $P_c(4312)$, $P_c(4440)$, and $P_c(4457)$ in the $J/\psi p$ invariant mass spectrum of the $\Lambda_b\to J/\psi p K$ \cite{Aaij:2019vzc} weak decay, which show a characteristic mass spectrum consistent with that of hidden-charm molecular pentaquark, which was predicted in Refs. \cite{Li:2014gra,Karliner:2015ina,Wu:2010jy,Wang:2011rga,Yang:2011wz,Wu:2012md,Chen:2015loa}.
Finally, it provides strong evidence to support the existence of the hidden-charm meson-baryon molecular states. Of course, there also exist other possible interpretations to the $P_c(4312)$, $P_c(4440)$, and $P_c(4457)$ \cite{Liu:2013waa,Hosaka:2016pey,Chen:2016qju,Richard:2016eis,Lebed:2016hpi,Brambilla:2019esw,Liu:2019zoy,Chen:2022asf,Olsen:2017bmm,Guo:2017jvc,Meng:2022ozq,Burns:2022uiv}.

Very recently, the LHCb Collaboration announced the observed of a $J/\psi\Lambda$ resonance in the $B^-\to J/\psi\Lambda\bar{p}$ process \cite{Collaboration:2022boa}, which has resonance parameters
\begin{eqnarray}
M=4338.2\pm0.7\pm0.4\,\mathrm{MeV},~~\Gamma=7.0\pm1.2\pm1.3\, \mathrm{MeV},\nonumber
\end{eqnarray}
and significance larger than $10\sigma$. Thus, this $J/\psi\Lambda$ resonance is refereed to be the $P_{\psi s}^\Lambda(4338)$, which is the candidate of hidden-charm pentaquark with strangeness, as predicted by former theoretical studies \cite{Chen:2016ryt,Wu:2010vk,Hofmann:2005sw,Anisovich:2015zqa,Wang:2015wsa,Feijoo:2015kts,Lu:2016roh,Xiao:2019gjd,Shen:2020gpw,Chen:2015sxa,Zhang:2020cdi,
Wang:2019nvm,Chen:2020uif,Peng:2020hql,Chen:2020opr,Liu:2020hcv,Dong:2021juy,Chen:2022onm,Chen:2020kco,Chen:2021cfl,Wang:2022neq,Chen:2021spf,Du:2021bgb,Hu:2021nvs,
Xiao:2021rgp,Zhu:2021lhd,Weng:2019ynv}. Before observing the $P_{\psi s}^\Lambda(4338)$, LHCb once reported an evidence of the enhancement structure ($P_{cs}(4459)$)  in the $J/\psi\Lambda$ invariant mass spectrum of the $\Xi_b^-\to J/\psi \Lambda K$ \cite{LHCb:2020jpq}, which is the candidate of hidden-charm pentaquark with strangeness \cite{Chen:2016ryt,Wu:2010vk,Hofmann:2005sw,Anisovich:2015zqa,Wang:2015wsa,Feijoo:2015kts,Lu:2016roh,Xiao:2019gjd,Shen:2020gpw,Chen:2015sxa,Zhang:2020cdi,
Wang:2019nvm,Chen:2020uif,Peng:2020hql,Chen:2020opr,Liu:2020hcv,Dong:2021juy,Chen:2022onm,Chen:2020kco,Chen:2021cfl,Wang:2022neq,Chen:2021spf,Du:2021bgb,Hu:2021nvs,
Xiao:2021rgp,Zhu:2021lhd,Weng:2019ynv}.

\begin{figure}[htbp]
\centering
\includegraphics[width=8.7cm]{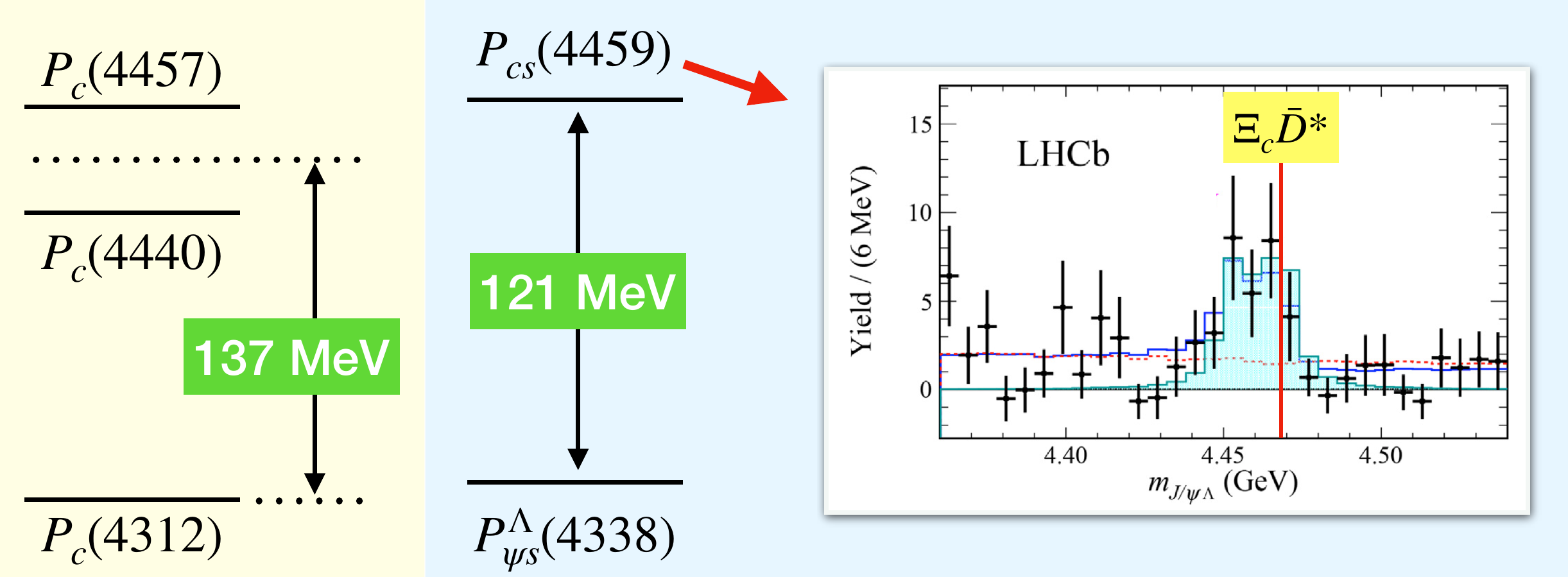}\\
\caption{Comparison of the mass spectrum of three $P_c$ states \cite{Aaij:2019vzc} and that of the $P_{\psi s}^\Lambda(4338)$ \cite{Collaboration:2022boa} and $P_{cs}(4459)$ \cite{LHCb:2020jpq}. Here, we also list the experimental data of the $P_{cs}(4459)$ from LHCb \cite{LHCb:2020jpq}.
}\label{fig1}
\end{figure}

In this work, we indicate that there exists a new characteristic mass spectrum applied to identify the molecular-type hidden-charm pentaquark with strangeness, which is inspired by the observed $P_{\psi s}^\Lambda(4338)$ and the reported $P_{cs}(4459)$. As shown in Fig. \ref{fig1}, the gap between the $P_c(4312)$ mass and the average mass of the $P_c(4440)$ and $P_c(4457)$ is 137 MeV, which is similar to the mass gap of the $P_{cs}(4459)$ and the $P_{\psi s}^\Lambda(4338)$.
{This is analogous to the similarity of the mass gaps for the $\omega$ and $\phi$ meson families \cite{Wang:2012wa}.}
We also notice an interesting phenomenon, and there exists corresponding relations of these $ P_c$ and $P_{cs}$ states, i.e., the $P_{\psi s}^\Lambda(4338)$ should correspond to the $P_c(4312)$, while the $P_{cs}(4459)$ structure corresponds to the $P_c(4440)$ and $P_c(4457)$. This fact makes us conjecture that the $P_{cs}(4459)$ enhancement structure should contain two substructures. If checking the LHCb data of the $J/\psi\Lambda$ invariant mass spectrum of the $\Xi_b^-\to J/\psi \Lambda K$ \cite{LHCb:2020jpq}, we find that there exists possible double peak structure slightly below the $\Xi_c\bar{D}^*$ threshold as indicated in Fig. \ref{fig1}, which just correspond to the $\Xi_c\bar{D}^*$ molecular states with $J^{P}=1/2^-$ and $3/2^-$. Such scenario can be tested in future experiment like the LHCb. As the partner of $P_c(4312)$, the $\Xi_c\bar{D}$ molecular state with $J^P=1/2^-$ was studied in this work. We should indicate the fact that the central value of the mass of the newly reported $P_{\psi s}^\Lambda(4338)$ is above the $\Xi_c \bar{D}$ threshold, which is the difficulty to directly assign the newly observed $P_{\psi s}^\Lambda(4338)$ as the $\Xi_c\bar{D}$ molecular state with $J^P=1/2^-$. How to solving this puzzling phenomenon is also an interesting topic. In the final section, we will address this point. Inspired by the established molecular-type characteristic mass spectrum of three $P_c$ states in 2019 \cite{Aaij:2019vzc} and the molecular-type characteristic mass spectrum of the $P_{\psi s}^\Lambda(4338)$ and the $P_{cs}(4459)$ enhancement structure found in this work, we further give a new characteristic mass spectrum of the $\Xi_c^\prime\bar{D}^{(*)}$ molecular states, which will be accessible at experiment.

This paper is organized as the follows. After introduction, the calculation of the interactions of these discussed $\Xi_c^{(\prime)} \bar{D}^{(*)}$ systems will be given by adopting the OBE model in Sec. \ref{sec2}. With this preparation, we discuss the characteristic mass spectra of the $\Xi_c^{(\prime)} \bar{D}^{(*)}$ molecular pentaquarks in Sec. \ref{sec3}. Finally, we will give the discussion and conclusion in Sec. \ref{sec4}.

\section{The OBE effective potentials of the $\Xi_c^{(\prime)} \bar{D}^{(*)}$ systems}\label{sec2}

For getting the information of the interactions of the $\Xi_c^{(\prime)}\bar{D}^{(*)}$ systems, the one-boson-exchange (OBE) model is adopted. In the OBE model, the interaction between two hadrons is a direct consequence of the exchange of the allowed light mesons \cite{Chen:2016qju,Liu:2019zoy}, which is one of the effective ways to investigate the interactions between hadrons. Such a formalism is a straightforward extension of the traditional meson exchange model involved in the nuclear force \cite{Yukawa:1935xg}. Up to now, the OBE model has successfully applied the exploration of a series of hadronic molecular states including the $P_c$ states and the $T_{cc}$ state \cite{Chen:2016qju,Liu:2019zoy}. We should indicate that it is not the only way to get the interactions between hadrons, since there also exist other approaches like the chiral effective field theory \cite{Meng:2022ozq}. For the chiral effective field theory, the contact term is usually introduced, which is different from the treatment of the OBE model on this point. In fact, this difference reflects the different treatment to the short distant interactions between hadrons under two approaches. Despite all this, the OBE model and the chiral effective field theory usually can reach the same conclusion for the same hadronic molecular states.

In general, there exists three typical steps for deducing the effective potentials of these discussed $\Xi_c^{(\prime)} \bar D_s^{(*)}$ systems within the OBE model. As the first step, we should write out the scattering amplitudes $\mathcal{M}(h_1h_2\to h_3h_4)$ of the scattering processes $h_1h_2\to h_3h_4$ by considering the effective Lagrangian approach.

According to the heavy quark spin symmetry, the hadrons containing single heavy quark with total spin $J_{\pm}=j_l\pm1/2$ (except $j_l=0$) come into doublets, which can be written as the super-fields when constructing the compact effective Lagrangians. {For the pseudoscalar anti-charmed meson $\bar D$ with $I(J^P)=1/2(0^-)$ and vector anti-charmed meson $\bar D^*$ with $I(J^P)=1/2(1^-)$, they are degenerated in the heavy quark spin symmetry,} which can be written as the super-field $H^{(\overline{Q})}_a=(\bar{D}^{*(\overline{Q})\mu}_a\gamma_{\mu}-\bar{D}^{(\overline{Q})}_a\gamma_5)\frac{(1-{v}\!\!\!\slash)}{2}$ \cite{Ding:2008gr}. In the heavy quark limit, the charm baryons can be classified in terms of the flavor symmetry of the diquark, the spin of the charm baryon in the $6_F$ flavor representation is either 1/2 or 3/2, while the spin of the charm baryon in the $\bar 3_F$ flavor representation is only $1/2$. Here, we need to mention that $\Xi_c$ with $J^P=1/2^+$ denotes the $S$-wave charm baryon in $\bar{3}_F$ flavor representation, while $\Xi_c^\prime$ with $J^P=1/2^+$ is the $S$-wave charmed baryon in $6_F$ flavor representation. The involved effective Lagrangians for depicting the heavy hadrons $\mathcal{B}_c/\bar{D}^{(*)}$ coupling with the light scalar, pesudoscalar, and vector mesons read as \cite{Wise:1992hn,Casalbuoni:1992gi,Casalbuoni:1996pg,Yan:1992gz,Bando:1987br,Harada:2003jx,Ding:2008gr,Chen:2017xat}
\begin{eqnarray}
\mathcal{L}_{\mathcal{B}_{\bar{3}}\mathcal{B}_{\bar{3}}\sigma} &=& l_B\langle \bar{\mathcal{B}}_{\bar{3}}\sigma\mathcal{B}_{\bar{3}}\rangle,\\
\mathcal{L}_{\mathcal{B}_{\bar{3}}\mathcal{B}_{\bar{3}}\mathbb{V}}&=&
\frac{1}{\sqrt{2}}\beta_Bg_V\langle\bar{\mathcal{B}}_{\bar{3}}v\cdot\mathbb{V}
\mathcal{B}_{\bar{3}}\rangle,\\
\mathcal{L}_{\mathcal{B}_{6}\mathcal{B}_{6}\sigma} &=&-l_S\langle\bar{\mathcal{B}}_6\sigma\mathcal{B}_6\rangle,\\
\mathcal{L}_{\mathcal{B}_6\mathcal{B}_6\mathbb{P}} &=&i\frac{g_1}{2f_{\pi}}\varepsilon^{\mu\nu\lambda\kappa}v_{\kappa}\langle\bar{\mathcal{B}}_6
\gamma_{\mu}\gamma_{\lambda}\partial_{\nu}\mathbb{P}\mathcal{B}_6\rangle,\\
\mathcal{L}_{\mathcal{B}_6\mathcal{B}_6\mathbb{V}}&=&
-\frac{\beta_Sg_V}{\sqrt{2}}\langle\bar{\mathcal{B}}_6v\cdot\mathbb{V}
\mathcal{B}_6\rangle\nonumber\\
    &&-i\frac{\lambda_S g_V}{3\sqrt{2}}\langle\bar{\mathcal{B}}_6\gamma_{\mu}\gamma_{\nu}
    \left(\partial^{\mu}\mathbb{V}^{\nu}-\partial^{\nu}\mathbb{V}^{\mu}\right)
    \mathcal{B}_6\rangle,\\
\mathcal{L}_{\mathcal{B}_{\bar{3}}\mathcal{B}_6\mathbb{P}} &=& -\sqrt{\frac{1}{3}}\frac{g_4}{f_{\pi}}\langle\bar{\mathcal{B}}_6\gamma^5\left(\gamma^{\mu}
+v^{\mu}\right)\partial_{\mu}\mathbb{P}\mathcal{B}_{\bar{3}}\rangle+h.c.,\\
\mathcal{L}_{\mathcal{B}_{\bar{3}}\mathcal{B}_6\mathbb{V}} &=&-\frac{\lambda_Ig_V}{\sqrt{6}}\varepsilon^{\mu\nu\lambda\kappa}v_{\mu}\langle \bar{\mathcal{B}}_6\gamma^5\gamma_{\nu}\left(\partial_{\lambda}\mathbb{V}_{\kappa}-\partial_{\kappa}\mathbb{V}_{\lambda}\right)\mathcal{B}_{\bar{3}}\rangle\nonumber\\
    &&+h.c.,\\
\mathcal{L}_{{\bar D}{\bar D}\sigma} &=&-2g_{\sigma} {\bar D}_{a} {\bar D}_a^{\dag}\sigma,\\
\mathcal{L}_{{\bar D}^{*}{\bar D}^{*}\sigma} &=&2g_{\sigma} {\bar D}_{a\mu}^* {\bar D}_a^{*\mu\dag}\sigma,\\
\mathcal{L}_{{\bar D}^{*}{\bar D}^{*}\mathbb{P}}&=&\frac{2ig}{f_{\pi}}v^{\alpha}\varepsilon_{\alpha\mu\nu\lambda}{\bar D}_a^{*\mu\dag}{\bar D}_b^{*\lambda}\partial^{\nu}{\mathbb{P}}_{ab},\\
\mathcal{L}_{{\bar D}{\bar D}\mathbb{V}}&=&\sqrt{2}\beta g_V {\bar D}_a {\bar D}_b^{\dag} v\cdot\mathbb{V}_{ab},\\
\mathcal{L}_{{\bar D}^{*}{\bar D}^{*}\mathbb{V}}&=&-\sqrt{2}\beta g_V {\bar D}_{a\mu}^* {\bar D}_b^{*\mu\dag}v\cdot\mathbb{V}_{ab}\nonumber\\
    &&-2\sqrt{2}i\lambda g_V {\bar D}_a^{*\mu\dag}{\bar D}_b^{*\nu}\left(\partial_{\mu}\mathbb{V}_{\nu}-\partial_{\nu}\mathbb{V}_{\mu}\right)_{ab}.
\end{eqnarray}
Here, $v_{\mu}=(1,\bm{0})$ is the four velocity under the non-relativistic approximation, and $\sigma$ stands for the scalar singlet meson. Additionally, the matrices $\mathcal{B}_{\bar{3}}$, $\mathcal{B}_6$, ${\mathbb{P}}$, and $\mathbb{V}_{\mu}$ have the standard forms as listed below
\begin{eqnarray}
\mathcal{B}_{\bar{3}} &=& \left(\begin{array}{ccc}
        0    &\Lambda_c^+      &\Xi_c^+\\
        -\Lambda_c^+       &0      &\Xi_c^0\\
        -\Xi_c^+      &-\Xi_c^0     &0
\end{array}\right),
\end{eqnarray}
\begin{eqnarray}
\mathcal{B}_6&=& \left(\begin{array}{ccc}
         \Sigma_c^{++}                  &\frac{\Sigma_c^{+}}{\sqrt{2}}     &\frac{\Xi_c^{(')+}}{\sqrt{2}}\\
         \frac{\Sigma_c^{+}}{\sqrt{2}}      &\Sigma_c^{0}    &\frac{\Xi_c^{(')0}}{\sqrt{2}}\\
         \frac{\Xi_c^{(')+}}{\sqrt{2}}    &\frac{\Xi_c^{(')0}}{\sqrt{2}}      &\Omega_c^{0}
\end{array}\right),
\end{eqnarray}
\begin{eqnarray}
{\mathbb{P}} &=& {\left(\begin{array}{cc}
       \frac{\pi^0}{\sqrt{2}}+\frac{\eta}{\sqrt{6}} &\pi^+ \\
       \pi^-       &-\frac{\pi^0}{\sqrt{2}}+\frac{\eta}{\sqrt{6}} \end{array}\right)},
       \end{eqnarray}
\begin{eqnarray}
{\mathbb{V}}_{\mu}&=& {\left(\begin{array}{cc}
       \frac{\rho^0}{\sqrt{2}}+\frac{\omega}{\sqrt{2}} &\rho^+ \\
       \rho^-       &-\frac{\rho^0}{\sqrt{2}}+\frac{\omega}{\sqrt{2}} \end{array}\right)}_{\mu},
\end{eqnarray}
respectively.

{As an effective way, the OBE model was widely used to investigate the interactions between hadrons. In calculation, the involved coupling constants are as input when discussing the problem of the hadronic molecular state. Usually, we prefer to fix the coupling constant if there exists the corresponding experimental data. For some coupling constants, we can estimate them by phenomenological models, where the relevant experimental information is absent.}
These coupling constants adopted in this work can be extracted from the experimental data or by the theoretical model, and the corresponding signs between these coupling constants can be fixed by the quark model \cite{Riska:2000gd}. Thus, we take $g_{\sigma}=0.76$, $l_B=-3.65$, $l_S=6.20$, $g=0.59$, $g_4=1.06$, $g_1=0.94$, $f_\pi=132~\rm{MeV}$, $\beta g_V=-5.25$, $\beta_B g_V=-6.00$, $\beta_S g_V=12.00$, $\lambda g_V =-3.27~\rm{GeV}^{-1}$, $\lambda_I g_V =-6.80~\rm{GeV}^{-1}$, and $\lambda_S g_V=19.20~\rm{GeV}^{-1}$ in the following numerical analysis \cite{Chen:2017xat,Chen:2019asm,Chen:2020kco}. Here, $g_{\sigma}$ can be deduced by the relation $g_{\sigma}=\widetilde g/2\sqrt{6}$ with $\widetilde g=3.73$ due to the spontaneously broken chiral symmetry \cite{Bardeen:2003kt}. The pionic coupling constant $g$ is determined by reproducing the experimental width of the $D^{*+} \to D^{0}\pi^{+}$ process \cite{Falk:1992cx,Isola:2003fh}. According to the vector meson dominance mechanism, $\beta$ and $g_V$ are fixed  \cite{Isola:2003fh,Cleven:2016qbn}. $\lambda$ can be obtained by comparing the form factor obtained from the lattice QCD with this calculated via the light cone sum rule \cite{Isola:2003fh}. $g_1$ is related to another coupling constant $g_4$ by the relation $g_1= \sqrt{8}g_4/3$ \cite{Liu:2011xc}, and $g_4$ can be extracted by the decay width of the $\Sigma_c^* \to \Lambda \pi$ process. $l_S$ is extracted with the chiral multiplet assumption \cite{Liu:2011xc}.
{For the coupling constants $l_B$, $\beta_B$, $\beta_S$, $\lambda_I$, and $\lambda_S$, we can calculate them by the quark model \cite{Riska:2000gd}, which is a popular approach to determine the coupling constants. Here, for the discussed vertex, we can write out its amplitude at hadronic and quark levels. By connecting these two amplitudes, finally the coupling constant can be obtained.}
{Here, we need to indicate that these coupling constants adopted in this work were applied to  reproduce the masses of the observed three $P_c$ states under the $S$-wave isoscalar $\Sigma_c \bar D^{(*)}$ molecular assignment \cite{Chen:2019asm}.}

With the above preparation, the obtained effective potentials in the momentum space can be related to the corresponding scattering amplitudes via the Breit approximation \cite{Breitapproximation}. Finally, the effective potentials in the coordinate space can be obtained by the Fourier transformation. Since the discussed hadrons are not pointlike particles, we should introduce the monopole type form factor $\mathcal{F}(q^2,m_E^2) = (\Lambda^2-m_E^2)/(\Lambda^2-q^2)$ \cite{Tornqvist:1993ng,Tornqvist:1993vu} in each interaction vertex, which can compensate the effect from the off-shell exchanged mesons and depict the structure effect of interaction vertex.

In order to obtain the concrete effective potentials for these discussed hidden-charm molecular pentaquark systems with strangeness, we need to construct their wave functions. For these discussed $\Xi_c^{(\prime)}\bar{D}^{(*)}$ systems, their spin-orbital wave functions can be expressed as
\begin{eqnarray}
|\Xi_c^{(\prime)}\bar{D}({}^{2S+1}L_{J})\rangle&=&\sum_{m,m_L}C^{J,M}_{\frac{1}{2}m,Lm_L}\chi_{\frac{1}{2}m}|Y_{L,m_L}\rangle,\\
|\Xi_c^{(\prime)}\bar{D}^*({}^{2S+1}L_{J})\rangle&=&\sum_{m,m',m_S,m_L}C^{S,m_S}_{\frac{1}{2}m,1m'}C^{J,M}_{Sm_S,Lm_L}\chi_{\frac{1}{2}m}\nonumber\\
&&\times\epsilon_{m'}^{\mu}|Y_{L,m_L}\rangle.
\end{eqnarray}
In the above expressions, the constant $C^{e,f}_{ab,cd}$ denotes the Clebsch-Gordan coefficient, and $|Y_{L,m_L}\rangle$ is the spherical harmonics function. The polarization vector $\epsilon_{m}^{\mu}\,(m=0,\,\pm1)$ with spin-1 field is written as $\epsilon_{\pm}^{\mu}= \left(0,\,\pm1,\,i,\,0\right)/\sqrt{2}$ and $\epsilon_{0}^{\mu}= \left(0,0,0,-1\right)$ in the static limit, and the $\chi_{\frac{1}{2}m}$ stands for the spin wave function for the charmed baryons $\Xi_c^{(\prime)}$. And then, we summarize the flavor wave functions $|I,I_{3}\rangle$ for these discussed $\Xi_c^{(\prime)}\bar{D}^{(*)}$ systems in Table \ref{flavor}.
\renewcommand\tabcolsep{0.62cm}
\renewcommand{\arraystretch}{1.50}
\begin{table}[!htpb]
\centering
\caption{Flavor wave functions for these discussed $\Xi_c^{(\prime)}\bar{D}^{(*)}$ systems. Here, $I$ and $I_3$ are their isospins and the third components, respectively. }\label{flavor}
{\begin{tabular}{c|lc}\toprule[1pt]\toprule[1pt]
$|I,I_3\rangle$    & Flavor wave functions \\\midrule[1.0pt]
$|1,1\rangle$ & $\left|\Xi_c^{(\prime)+}\bar{D}^{(*)0}\right\rangle$\\
$|1,0\rangle$ & $\sqrt{\frac{1}{2}}\left|\Xi_c^{(\prime)+}{D}^{(*)-}\right\rangle+\sqrt{\frac{1}{2}}\left|\Xi_c^{(\prime)0}\bar{D}^{(*)0}\right\rangle$\\
$|1,-1\rangle$ & $\left|\Xi_c^{(\prime)0}{D}^{(*)-}\right\rangle$\\
$|0,0\rangle$ & $\sqrt{\frac{1}{2}}\left|\Xi_c^{(\prime)+}{D}^{(*)-}\right\rangle-\sqrt{\frac{1}{2}}\left|\Xi_c^{(\prime)0}\bar{D}^{(*)0}\right\rangle$\\
\bottomrule[1pt]\bottomrule[1pt]
\end{tabular}}
\end{table}

With the standard procedures of the OBE model, the expressions of the effective potentials in the coordinate space for these investigated isoscalar  $\Xi_c^{(\prime)} \bar D_s^{(*)}$ systems are given by
\begin{eqnarray*}
\mathcal{V}^{\Xi_{c}\bar D}&=&2l_{B}g_{\sigma}Y_\sigma-\frac{3\beta \beta_B g_{V}^2}{4}Y_{\rho}+\frac{\beta \beta_B g_{V}^2}{4}Y_{\omega},\\
\mathcal{V}^{\Xi_{c}^{\prime}\bar D}&=&-l_Sg_{\sigma}Y_\sigma+\frac{3\beta \beta_S g_{V}^2}{8}Y_{\rho}-\frac{\beta \beta_S g_{V}^2}{8}Y_{\omega},\\
\mathcal{V}^{\Xi_{c}\bar D^*}&=&2l_{B}g_{\sigma}\mathcal{A}_{1}Y_\sigma-\frac{3\beta \beta_B g_{V}^2}{4}\mathcal{A}_{1}Y_{\rho}+\frac{\beta \beta_B g_{V}^2}{4}\mathcal{A}_{1}Y_{\omega},\\
\mathcal{V}^{\Xi_{c}^{\prime}\bar D^*}&=&-l_Sg_{\sigma}\mathcal{A}_{1}Y_\sigma-\frac{g_1 g}{4f_\pi^2}\left[\mathcal{A}_{2}\mathcal{O}_r+\mathcal{A}_{3}\mathcal{P}_r\right]Y_{\pi}\nonumber\\
&-&\frac{g_1 g}{36f_\pi^2}\left[\mathcal{A}_{2}\mathcal{O}_r+\mathcal{A}_{3}\mathcal{P}_r\right]Y_{\eta}\nonumber\\
&+&\frac{3\beta \beta_S g_{V}^2}{8}\mathcal{A}_{1}Y_{\rho}+\frac{3\lambda \lambda_S g_V^2}{18}\left[2\mathcal{A}_{2}\mathcal{O}_r-\mathcal{A}_{3}\mathcal{P}_r\right]Y_{\rho}\nonumber\\
&-&\frac{\beta \beta_S g_{V}^2}{8}\mathcal{A}_{1}Y_{\omega}-\frac{\lambda \lambda_S g_V^2}{18}\left[2\mathcal{A}_{2}\mathcal{O}_r-\mathcal{A}_{3}\mathcal{P}_r\right]Y_{\omega}.\\
\end{eqnarray*}
Here, $\mathcal{O}_r = \frac{1}{r^2}\frac{\partial}{\partial r}r^2\frac{\partial}{\partial r}$ and $\mathcal{P}_r = r\frac{\partial}{\partial r}\frac{1}{r}\frac{\partial}{\partial r}$. The function $Y_i$ is defined as
\begin{eqnarray}
Y_i\equiv \dfrac{e^{-m_ir}-e^{-\Lambda_i r}}{4\pi r}-\dfrac{\Lambda_i^2-m_i^2}{8\pi\Lambda_i}e^{-\Lambda_i r}.
\end{eqnarray}
Additionally, we also define several operators, which include $\mathcal{A}_{1}=\chi^{\dagger}_3\left({\bm\epsilon^{\dagger}_{4}}\cdot{\bm\epsilon_{2}}\right)\chi_1$, $\mathcal{A}_{2}=\chi^{\dagger}_3\left[{\bm\sigma}\cdot\left(i{\bm\epsilon_{2}}\times{\bm\epsilon^{\dagger}_{4}}\right)\right]\chi_1$, and
$\mathcal{A}_{3}=\chi^{\dagger}_3T({\bm\sigma},i{\bm\epsilon_{2}}\times{\bm\epsilon^{\dagger}_{4}},\hat{\bm r})\chi_1$. Here, the tensor force operator is expressed as $T({\bm x},{\bm y},\hat{\bm r})= 3\left(\hat{\bm r} \cdot {\bm x}\right)\left(\hat{\bm r} \cdot {\bm y}\right)-{\bm x} \cdot {\bm y}$ with $\hat{\bm r}={\bm r}/|{\bm r}|$. In the effective potentials, the corresponding matrix elements $\langle f|\mathcal{A}_k|i\rangle$ can be obtained by these operators $\mathcal{A}_k$ sandwiched  between the relevant spin-orbit wave functions of the initial and final states. For example, the matrices element $\langle\Xi_c^{(\prime)}\bar{D}^{*}|\mathcal{A}_{1}|\Xi_c^{(\prime)}\bar{D}^{*}\rangle$ with $J=1/2$ can be expressed as ${\left(\begin{array}{cc}
\mathcal{A}_{1}^{11}&\mathcal{A}_{1}^{12}\\ \mathcal{A}_{1}^{21}&\mathcal{A}_{1}^{22}\\\end{array}\right)}$ with
\begin{eqnarray}
\mathcal{A}_{1}^{11}&=&\langle\Xi_c^{(\prime)}\bar{D}^{*}({}^2\mathbb{S}_{\frac{1}{2}})|\mathcal{A}_{1}|\Xi_c^{(\prime)}\bar{D}^{*}({}^2\mathbb{S}_{\frac{1}{2}})\rangle,\nonumber\\
\mathcal{A}_{1}^{12}&=&\langle\Xi_c^{(\prime)}\bar{D}^{*}({}^4\mathbb{D}_{\frac{1}{2}})|\mathcal{A}_{1}|\Xi_c^{(\prime)}\bar{D}^{*}({}^2\mathbb{S}_{\frac{1}{2}})\rangle,\nonumber\\
\mathcal{A}_{1}^{21}&=&\langle\Xi_c^{(\prime)}\bar{D}^{*}({}^2\mathbb{S}_{\frac{1}{2}})|\mathcal{A}_{1}|\Xi_c^{(\prime)}\bar{D}^{*}({}^4\mathbb{D}_{\frac{1}{2}})\rangle,\nonumber\\
\mathcal{A}_{1}^{22}&=&\langle\Xi_c^{(\prime)}\bar{D}^{*}({}^4\mathbb{D}_{\frac{1}{2}})|\mathcal{A}_{1}|\Xi_c^{(\prime)}\bar{D}^{*}({}^4\mathbb{D}_{\frac{1}{2}})\rangle.\nonumber
\end{eqnarray}
In Table~\ref{matrix}, we collect the obtained relevant operator matrix elements $\langle f|\mathcal{A}_k|i\rangle$, which are used in our calculation.
\renewcommand\tabcolsep{0.22cm}
\renewcommand{\arraystretch}{1.50}
\begin{table}[htbp]
  \caption{The relevant operator matrix elements $\langle f|\mathcal{A}_k|i\rangle$, which are obtained by sandwiching these operators between the relevant spin-orbit wave functions.}\label{matrix}
  \begin{tabular}{c|cc}\toprule[1pt]\toprule[1pt]
   {{{Spin}}} & $J=1/2$     & $J=3/2$  \\\midrule[1pt]
 $\langle\Xi_c^{(\prime)}\bar{D}^{*}|\mathcal{A}_{1}|\Xi_c^{(\prime)}\bar{D}^{*}\rangle$
            &$\left(\begin{array}{cc} 1 & 0 \\ 0 & 1\end{array}\right)$ &$\left(\begin{array}{ccc} 1 & 0& 0 \\ 0 & 1& 0 \\ 0& 0& 1 \end{array}\right)$\\
 $\langle\Xi_c^{\prime}\bar{D}^{*}|\mathcal{A}_{2}|\Xi_c^{\prime}\bar{D}^{*}\rangle$
 &$\left(\begin{array}{cc} -2 & 0 \\ 0 & 1\end{array}\right)$ &$\left(\begin{array}{ccc} 1 & 0& 0 \\ 0 & -2& 0 \\ 0& 0& 1 \end{array}\right)$\\
  $\langle\Xi_c^{\prime}\bar{D}^{*}|\mathcal{A}_{3}|\Xi_c^{\prime}\bar{D}^{*}\rangle$
           &$\left(\begin{array}{cc} 0 & -\sqrt{2} \\ -\sqrt{2} & -2\end{array}\right)$&$\left(\begin{array}{ccc} 0 & 1& 2 \\ 1 & 0& -1 \\ 2 & -1& 0 \end{array}\right)$\\
\bottomrule[1pt]\bottomrule[1pt]
\end{tabular}
\end{table}

When discussing the bound state properties of the $S$-wave isoscalar $\Xi_{c}\bar D^*$ states with $J^P=1/2^-$ and $3/2^-$ after including the coupled channels $\Xi_{c}\bar D^*$ and $\Xi_{c}^{\prime}\bar D^*$, we need the effective potentials in the coordinate space for the $\Xi_{c}\bar D^* \to \Xi_{c}^{\prime}\bar D^*$ process, i.e,
\begin{eqnarray*}
\mathcal{V}^{\Xi_{c}\bar D^*\to \Xi_{c}^{\prime}\bar D^*}&=&-\frac{gg_4}{2\sqrt{6}f_\pi^2}\left[\mathcal{A}_{2}\mathcal{O}_r+\mathcal{A}_{3}\mathcal{P}_r\right]Y_{\pi0}\nonumber\\
&+&\frac{gg_4}{6\sqrt{6}f_\pi^2}\left[\mathcal{A}_{2}\mathcal{O}_r+\mathcal{A}_{3}\mathcal{P}_r\right]Y_{\eta0}\nonumber\\
&-&\frac{\lambda \lambda_I g_V^2}{\sqrt{6}}\left[2\mathcal{A}_{2}\mathcal{O}_r-\mathcal{A}_{3}\mathcal{P}_r\right]Y_{\rho0}\nonumber\\
&+&\frac{\lambda \lambda_I g_V^2}{3\sqrt{6}}\left[2\mathcal{A}_{2}\mathcal{O}_r-\mathcal{A}_{3}\mathcal{P}_r\right]Y_{\omega0}.
\end{eqnarray*}
Here, $m_0=\sqrt{m^2-q_0^2}$ and $\Lambda_0=\sqrt{\Lambda^2-q_0^2}$ with $q_0=0.06\,{\rm {GeV}}$.

\section{The characteristic mass spectra of the $\Xi_c^{(\prime)} \bar{D}^{(*)}$ molecular pentaquarks}\label{sec3}

Based on the obtained effective potentials in the coordinate space for these discussed hidden-charm pentaquarks with strangeness, we can obtain the bound state solutions by solving the coupled channel Schr$\rm{\ddot{o}}$dinger equation, where the obtained bound state solutions mainly include the binding energy $E$, the root-mean-square radius $r_{\rm RMS}$, and the probabilities for different components $P_i$, which may provide us the critical information to judge whether these discussed hidden-charm molecular pentaquark states with strangeness exist or not. {In the coupled channel analysis, the probabilities for different components $P_i$ are useful information to reflect the properties of the hadronic molecular states, except for the binding energy $E$ and the root-mean-square radius $r_{\rm RMS}$. If the coupled system is not dominant by the lowest mass threshold channel among the selected coupled channels, where usually the obtained root-mean-square radius $r_{{\rm RMS}}$ is too small, this coupled system is not recommended to be a hadronic molecular state (see Ref. \cite{Chen:2017xat} for more details).}

Of course, when judging whether the loosely bound state is the possible hadronic molecular candidate, we also need to specify two points: (1) {As a free parameter, the cutoff value of the form factor  cannot be determined exactly due to the lack of relevant experiment data. In realistic calculation, one should find bound state solution by changing the cutoff value. According to the experience of studying the deuteron by the OBE model, the cutoff value should not be far away from 1 GeV \cite{Machleidt:1987hj,Epelbaum:2008ga,Esposito:2014rxa,Chen:2016qju}. Furthermore, we can reproduce the masses of the observed three $P_c$ states \cite{Chen:2019asm} and $T_{cc}$ state \cite{Li:2021zbw} under the hadronic molecular picture when taking the cutoff values around 1 GeV. In the present work, we still take the cutoff range around 1 GeV to discuss the bound state properties of the $S$-wave isoscalar $\Xi_c^{(\prime)}\bar D^{(*)}$ systems}; (2) The reasonable binding energy for the possible hadronic molecular candidate should be at most tens of MeV, and the corresponding typical size should be larger than the size of all the included component hadrons \cite{Chen:2017xat}. These criteria may provide useful hints to identify the hidden-charm molecular pentaquark candidates with strangeness.

In our calculation, the masses of these involved hadrons are $m_\sigma=600.00$ MeV, $m_\pi=137.27$ MeV, $m_\eta=547.86$ MeV, $m_\rho=775.26$ , $m_\omega=782.66$ MeV, $m_{D}=1867.25$ MeV, $m_{D^*}=2008.56$ MeV, $m_{\Xi_{c}}=2469.08$ MeV, and $m_{\Xi_{c}^{\prime}}=2578.45$ MeV, which are taken from the Particle Data Group \cite{ParticleDataGroup:2022pth}.

\subsection{The characteristic mass spectrum of the $\Xi_c\bar{D}^{(*)}$ molecular pentaquarks}

{For the $S$-wave isoscalar $\Xi_{c}\bar D^{(*)}$ systems, we list the obtained bound state solutions by considering the $S$-$D$ wave mixing effect in Table. \ref{XicDSD}. For the $S$-wave isoscalar $\Xi_{c}\bar D$ state with $J^P=1/2^-$, we obtain the bound state solution by setting the cutoff parameter $\Lambda$ around 1.41 GeV. For the $S$-wave isoscalar $\Xi_{c}\bar D^*$ states with $J^P=1/2^-$ and $3/2^-$, there exist the bound state solutions when the cutoff parameter is taken to be around 1.39 GeV. Here, the probabilities for the $D$-wave components are zero for these two states, since the contribution of the tensor forces from the $S$-$D$ wave mixing effect disappears for the case of the isoscalar $\Xi_{c}\bar D^*$ interactions. In addition, there exists mass degeneration for the $S$-wave isoscalar $\Xi_{c}\bar D^*$ states with $J^P=1/2^-$ and $3/2^-$ when adopting same cutoff value in the $S$-$D$ wave mixing analysis.
\renewcommand\tabcolsep{0.45cm}
\renewcommand{\arraystretch}{1.50}
\begin{table}[!htbp]
\centering
\caption{Bound state properties for the $S$-wave isoscalar $\Xi_{c}\bar D^{(*)}$ systems by considering the $S$-$D$ wave mixing effect. Here, the cutoff $\Lambda$, binding energy $E$, and root-mean-square radius $r_{{\rm RMS}}$ are in units of $ \rm{GeV}$, $\rm {MeV}$, and $\rm {fm}$, respectively.}\label{XicDSD}
\begin{tabular}{cccc}\toprule[1.0pt]\toprule[1.0pt]
\multicolumn{4}{c}{$\Xi_{c}\bar D(J^P=1/2^-)$}\\\midrule[1.0pt]
$\Lambda$  &$E$ &$r_{\rm RMS}$ \\\hline
1.41&$-0.35$ &4.73     \\
1.61&$-4.82$ &1.64      \\
1.79&$-12.49$ &1.10     \\\midrule[1.0pt]
\multicolumn{4}{c}{$\Xi_{c}\bar D^{*}(J^P=1/2^-)$}\\\midrule[1.0pt]
$\Lambda$  &$E$ &$r_{\rm RMS}$ &P(${}^2\mathbb{S}_{\frac{1}{2}}/{}^4\mathbb{D}_{\frac{1}{2}}$)\\\hline
1.39&$-0.34$ &4.70&\textbf{100.00}/$o(0)$     \\
1.57&$-4.71$ &1.63&\textbf{100.00}/$o(0)$      \\
1.74&$-12.21$ &1.10&\textbf{100.00}/$o(0)$     \\\midrule[1.0pt]
\multicolumn{4}{c}{$\Xi_{c}\bar D^{*}(J^P=3/2^-)$}\\\midrule[1.0pt]
$\Lambda$ &$E$ &$r_{\rm RMS}$ &P(${}^4\mathbb{S}_{\frac{3}{2}}/{}^2\mathbb{D}_{\frac{3}{2}}/{}^4\mathbb{D}_{\frac{3}{2}}$)\\\hline
1.39&$-0.34$ &4.70&\textbf{100.00}/$o(0)$/$o(0)$\\
1.57&$-4.71$ &1.63&\textbf{100.00}/$o(0)$/$o(0)$\\
1.74&$-12.21$ &1.10&\textbf{100.00}/$o(0)$/$o(0)$\\
\bottomrule[1.0pt]\bottomrule[1.0pt]
\end{tabular}
\end{table}}

In the above discussions, we indicate that these discussed $S$-wave isoscalar $\Xi_{c}\bar D^{(*)}$ systems can be viewed as the hidden-charm molecular pentaquark candidates with single strangeness within the OBE model by considering the $S$-$D$ wave mixing effect. However, there exist mass degeneration for the $S$-wave isoscalar $\Xi_{c}\bar D^*$ states with $J^P=1/2^-$ and $3/2^-$ when adopting same cutoff value in the $S$-$D$ wave mixing analysis. Since the observation of the charmoniumlike state $X(3872)$, various corrections \cite{Chen:2016qju} including the coupled channel effect \cite{Li:2012bt,Li:2012cs,Li:2012ss} have been introduced to discuss the fine properties of the hadron-hadron interactions in the OBE model, which may decorate the bound state properties of hadronic molecular states \cite{Li:2012bt,Li:2012cs,Li:2012ss,Chen:2015add,Chen:2017jjn,Chen:2017xat,Chen:2018pzd,Wang:2019nwt,Chen:2019asm,Wang:2019aoc,Du:2019pij,Yamaguchi:2019seo,Wang:2020dya,Wang:2020bjt,Chen:2020kco,Wang:2021ajy,Chen:2021kad,Wang:2021hql,Wang:2021yld,Li:2021zbw,Yang:2021sue,Du:2021fmf,Chen:2022onm,Burns:2022uiv,Roca:2006sz,Oset:2010tof,Liu:2011xc,Xie:2012np,Xiao:2013yca,Xiao:2013jla,Ozpineci:2013zas,Liang:2014eba,Roca:2015dva,Moir:2016srx,Kang:2016zmv,Albaladejo:2016eps,Shimizu:2016rrd,Shimizu:2017xrg,Yamaguchi:2016ote,Lu:2016kxm,Shimizu:2018ran,Pavao:2018xub,Sekihara:2018tsb,Valderrama:2018bmv,Dias:2018qhp,He:2018plt,He:2017lhy,Shen:2017ayv,Montana:2017kjw,Meng:2019ilv,Bruns:2019fwi,Yang:2018amd,Yu:2019yfr,Burns:2019iih,Xiao:2019gjd,Xiao:2019aya,Yu:2018yxl,Wang:2019niy,Ortega:2018cnm,Zhu:2020vto,Yang:2019rgw,Chen:2020yvq,Chen:2021vhg,Ikeno:2020mra,Zhu:2021lhd,Yalikun:2021bfm,Peng:2020hql,Hao:2022vwt,Ikeno:2021mcb,Kamiya:2022thy,Dai:2022ulk,Zhu:2022fyb,Wang:2022oof,Yalikun:2021dpk,Albaladejo:2021vln,Wang:2021lwy}.

For the $\Xi_c\bar D^{(*)}$ molecular systems, the authors of Refs. \cite{Chen:2020kco,Chen:2022onm} once presented a coupled channel analysis within the OBE model. Here, for the $S$-wave isoscalar $\Xi_{c}\bar D^*$ state with $J^P=3/2^-$, the coupled channel effect is helpful to form this molecular state \cite{Chen:2020kco}, since the cutoff value becomes smaller after including the coupled channel effect compared with the situation without considering the coupled channel effect. However, when discussing the isoscalar $\Xi_{c}\bar D^*$ state with $J^P=1/2^-$ with a coupled channel analysis, a puzzling phenomenon appears \cite{Chen:2020kco}, where a higher $\Xi_{c}^{\prime}\bar D^*$ channel among these selected coupled channels becomes dominant. Obviously, a jump occurs which is not reasonable. The reason resulting in this puzzling phenomenon is that the author of Ref.  \cite{Chen:2020kco} took the same cutoff value for these considered channels in their coupled channel analysis. Obviously, this treatment should be improved to ensure that the coupled channel effect only plays the role to decorate the discussed pure state, which was not realised in these early studies \cite{Chen:2022onm,Chen:2021kad,Wang:2021hql,Chen:2020kco,Wang:2019aoc,Chen:2017xat,Wang:2021yld,Wang:2020dya}.

An approaches to overcome such problem is proposed in this work. In practice, the cutoff values for the involved coupled channels can be different. Along this line, we discuss the bound state properties of the isoscalar $\Xi_{c}\bar D^*$ molecular states with $J^P=1/2^-$ and $3/2^-$, where we adopt different cutoff values for the $\Xi_{c}\bar D^*$ and $\Xi_{c}^{\prime}\bar D^*$ channels.
{In Table \ref{XicDCC}, we present the bound state solutions of the $S$-wave isoscalar $\Xi_{c}\bar D^*$ states with $J^P=1/2^-$ and $3/2^-$ by the coupled channel analysis. The above fact shows that the coupled channel effect indeed may decorate the bound state properties of the $S$-wave isoscalar $\Xi_{c}\bar D^*$ state with $J^P=1/2^-$, and ensure that the jump phenomenon mentioned above does not happen. Thus, the conclusion of the existence of the $S$-wave isoscalar $\Xi_{c}\bar D^*$ molecular state with $J^P=1/2^-$  does not change when considering the coupled channel effect. We should indicate that the former studies in Refs. \cite{Zhu:2021lhd,Wang:2019nvm,Chen:2021cfl} also support the existence of the $S$-wave isoscalar $\Xi_{c}\bar D^*$ molecular states with $J^P=1/2^-$ and $3/2^-$. In this work, we can find the bound state solutions for the $S$-wave isoscalar $\Xi_{c}\bar D^*$ states with $J^P=1/2^-$ and $3/2^-$, where the $\Xi_{c}\bar D^*$ is the dominant channel. Different from the case of the single channel analysis, there exists the mass difference for the $S$-wave isoscalar $\Xi_{c}\bar D^*$ states with $J^P=1/2^-$ and $3/2^-$ when considering the coupled channel effect.}

\renewcommand\tabcolsep{0.35cm}
\renewcommand{\arraystretch}{1.50}
\begin{table}[!htbp]
\centering
\caption{Bound state properties for the $S$-wave isoscalar $\Xi_{c}\bar D^{*}$ system by performing the coupled channel analysis. Here, the cutoff $\Lambda$, binding energy $E$, and root-mean-square radius $r_{{\rm RMS}}$ are in units of $ \rm{GeV}$, $\rm {MeV}$, and $\rm {fm}$, respectively. Additionally, $\Lambda$ and $\Lambda^{\prime}$ denote the cutoff parameters of the $\Xi_{c}\bar D^*$ and $\Xi_{c}^{\prime}\bar D^*$ channels, respectively.}\label{XicDCC}
\begin{tabular}{ccccc}\toprule[1.0pt]\toprule[1.0pt]
$\Lambda$ &$\Lambda^{\prime}$ &$E$ &$r_{\rm RMS}$ &P($\Xi_{c}\bar D^*/\Xi_{c}^{\prime}\bar D^*$)\\\midrule[1.0pt]
\multicolumn{5}{c}{$\Xi_{c}\bar D^{*}(J^P=1/2^-)$}\\\midrule[1.0pt]
1.12&0.92&$-0.30$ &4.74&\textbf{97.75}/2.25 \\
1.16&0.96&$-4.33$ &1.58&\textbf{89.46}/10.54 \\
1.20&1.00&$-14.67$ &0.89&\textbf{77.76}/22.24 \\\midrule[1.0pt]
\multicolumn{5}{c}{$\Xi_{c}\bar D^{*}(J^P=3/2^-)$}\\\midrule[1.0pt]
1.31&1.11&$-0.29$ &4.87&\textbf{99.73}/0.27\\
1.43&1.23&$-4.52$ &1.64&\textbf{98.54}/1.46\\
1.56&1.36&$-15.01$ &0.98&\textbf{96.48}/3.52\\
\bottomrule[1.0pt]\bottomrule[1.0pt]
\end{tabular}
\end{table}

We also need to mention that the $P_{cs}(4459)$ existing in the $J/\psi \Lambda$ invariant mass spectrum may be described by two peak structures with the masses of 4454.9 MeV and 4467.9 MeV \cite{LHCb:2020jpq}. According to Table \ref{XicDCC}, the masses of the $S$-wave isoscalar $\Xi_c\bar D^{*}$ state with $J^P=1/2^-$ may be around 4454.9 MeV, we can reproduce this mass with the cutoff parameters of the $\Xi_{c}\bar D^*$ and $\Xi_{c}^{\prime}\bar D^*$ channels respectively around 1.22 GeV and 1.02 GeV, which are close to the reasonable range around 1.00 GeV \cite{Tornqvist:1993ng,Tornqvist:1993vu,Wang:2019nwt,Chen:2017jjn}. Here, its binding energy and root-mean-square radius are $-22.70$ MeV and 0.73 fm, and the probabilities of the $\Xi_{c}\bar D^*$ and $\Xi_{c}^{\prime}\bar D^*$ components are around 72\% and 28\%, respectively.

The corresponding thresholds of the $\Xi_{c}\bar D$ and $\Xi_{c}\bar D^*$ channels are 4336.33 MeV and 4477.64 MeV, so the masses of the $S$-wave isoscalar $\Xi_{c}\bar D^*$ states are larger than that of the $S$-wave isoscalar $\Xi_{c}\bar D$ state. Based on the analysis mentioned above, it is clear that the masses of these discussed hidden-charm molecular pentaquarks with strangeness satisfy the relation $m[\Xi_{c}\bar D]<m[\Xi_{c}\bar D^*]$. Generally, the above analysis presents a characteristic mass spectrum of hidden-charm pentaquark with strangeness. {Additionally, our obtained $S$-wave isoscalar $\Xi_c\bar{D}$ state with $J^P=1/2^-$ has the mass lower than that of the observed $P_{\psi s}^\Lambda(4338)$ \cite{Collaboration:2022boa}. However, the $P_{\psi s}^\Lambda(4338)$ was reported in the $J/\psi \Lambda$ invariant mass spectrum via the $B^-\to J/\psi \Lambda \bar{p}$ process with the width about 7 MeV. Thus, there is still some possibility that the origin of the $P_{\psi s}^\Lambda(4338)$ can be due to the $S$-wave isoscalar $\Xi_c\bar D$ molecular state with $J^P=1/2^-$, which should be checked in future experiment. Especially, confirming it via the $\Xi_b^-\to J/\psi \Lambda K$ process is strongly suggested here.}

{In Ref. \cite{Guo:2019kdc}, the authors once estimated the probabilities of finding the $\Sigma_c \bar D^{(*)}$ and $J/\psi p$ components inside the $P_c$ states within the effective range expansion and the resonance compositeness relations, and concluded that the weight of the $\Sigma_c \bar D^{(*)}$ component around 99\% is much larger than that of the $J/\psi p$ component. Thus, most of the previous theoretical studies usually ignored the contribution of the $J/\psi p$ channel when explaining the observed $P_c$ states as the $S$-wave isoscalar $\Sigma_c \bar D^{(*)}$ molecular states. Here, we would like to mention that there exists similar spectroscopy behavior for the $S$-wave isoscalar $\Sigma_c \bar D^{(*)}$ and $\Xi_c \bar D^{(*)}$ molecular states, which reflects that the weight of the $\Xi_c \bar D^{(*)}$ component should be much larger than that of the $J/\psi \Lambda$ component for these discussed hidden-charm molecular pentaquarks with single strangeness. In the present work, we do not consider the contribution of the $J/\psi \Lambda$ channel when discussing the bound state properties of the $S$-wave isoscalar $\Xi_c^{(\prime)}\bar D^{(*)}$ states.}

\subsection{The characteristic mass spectrum of the $\Xi_c^{\prime}\bar{D}^{(*)}$ molecular pentaquarks}

\begin{figure}[htbp]
\centering
\includegraphics[width=8.7cm]{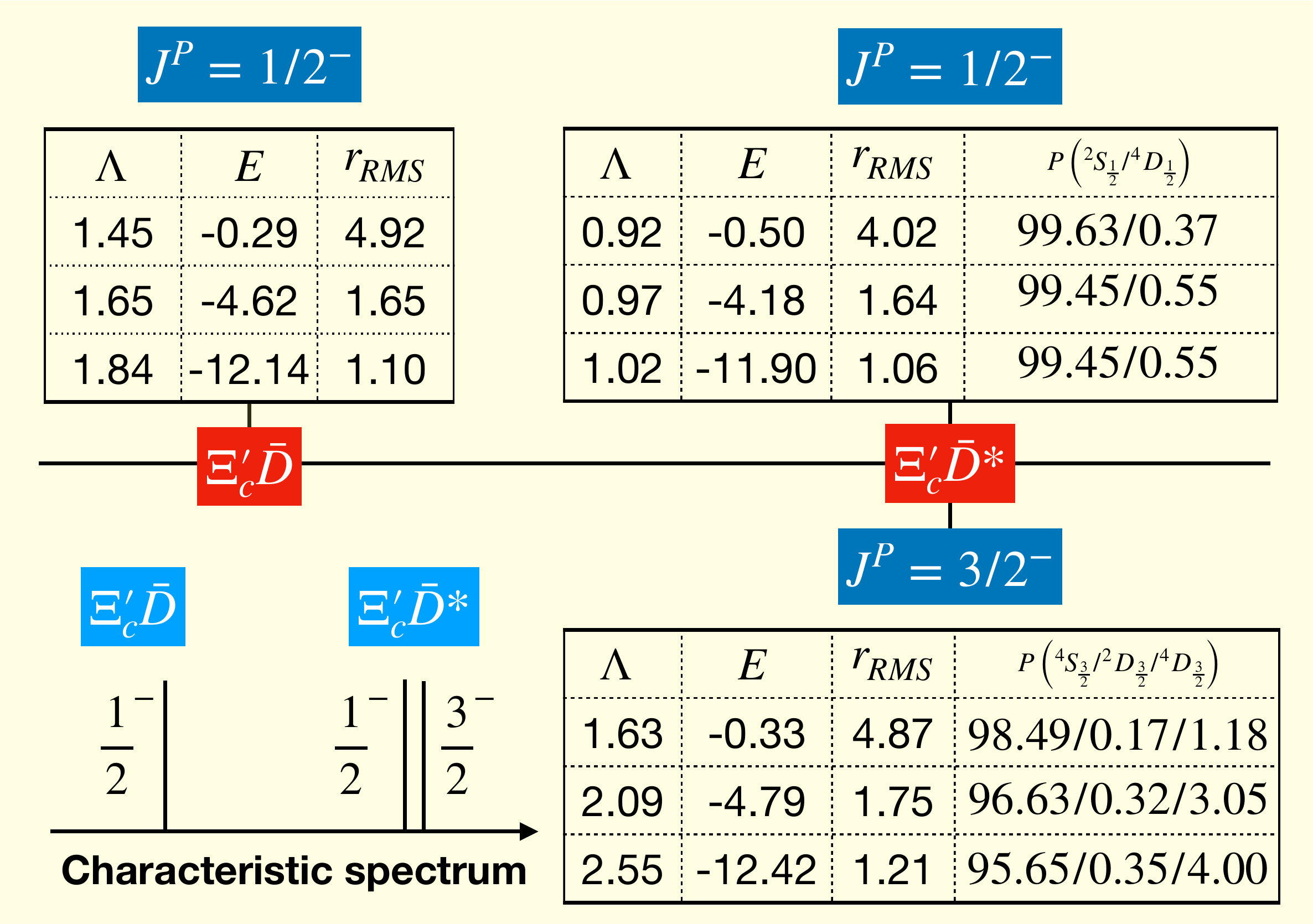}\\
\caption{Bound state properties for the $S$-wave isoscalar $\Xi_{c}^{\prime}\bar D^{(*)}$ systems. Here, the cutoff $\Lambda$, binding energy $E$, and root-mean-square radius $r_{{\rm RMS}}$ are in units of $ \rm{GeV}$, $\rm {MeV}$, and $\rm {fm}$, respectively.
}\label{fig3}
\end{figure}

Following the procedure discussed above, we present the binding energy, root-mean-square radius, and probabilities for different components for the $S$-wave isoscalar $\Xi_{c}^{\prime}\bar D^{(*)}$ systems in Fig. \ref{fig3}. For the $S$-wave isoscalar $\Xi_{c}^{\prime} \bar D$ state with $J^P=1/2^-$, the bound state solution can be found when the cutoff parameter is fixed to be larger than 1.45 GeV. For the $S$-wave isoscalar $\Xi_{c}^{\prime} \bar D^*$ system, the $\pi$, $\sigma$, $\eta$, $\rho$, and $\omega$ exchanges contribute to the total effective potentials. For the $S$-wave isoscalar $\Xi_{c}^{\prime}\bar D^*$ state with $J^P=1/2^-$, there exists the bound state solutions with the cutoff parameter around 0.92 GeV, while we can obtain loosely bound state solutions for the $S$-wave isoscalar $\Xi_{c}^{\prime}\bar D^*$ state with $J^P=3/2^-$ when we tune the cutoff parameter to be around 1.63 GeV. Thus, we can conclude these discussed states are the isoscalar hidden-charm molecular pentaquark candidates with strangeness, which is consistent with the conclusions in Refs. \cite{Zhu:2021lhd,Wang:2019nvm,Chen:2021cfl,Chen:2022onm}. Additionally, the largest mass of these isoscalar hidden-charm molecular pentaquark candidates with strangeness is the $S$-wave isoscalar $\Xi_{c}^{\prime}\bar D^*$ state with $J^P=3/2^-$, followed by the the $S$-wave isoscalar $\Xi_{c}^{\prime}\bar D^*$ state with $J^P=1/2^-$ and the $S$-wave isoscalar $\Xi_{c}^{\prime}\bar D$ state with $J^P=1/2^-$. Here, we need to point out that  the masses of the $S$-wave isoscalar $\Xi_{c}^{\prime}\bar D^*$ state with $J^P=3/2^-$ and $1/2^-$ are 4582.3 MeV and 4568.7 MeV based on the chiral effective field theory \cite{Wang:2019nvm}, and the masses of the $S$-wave isoscalar $\Xi_{c}^{\prime}\bar D^*$ state with $J^P=3/2^-$ and $1/2^-$ are 4582.1 MeV and 4564.9 MeV with a quark level interaction \cite{Chen:2021cfl}, which are comparable with our results.
For the $\Xi_c^\prime\bar{D}^{(*)}$ pentaquark systems, there also exists a characteristic mass spectrum as predicted above (see Fig. \ref{fig3}).

{For the $S$-wave loosely bound state, its bound state properties have close relation to the effective potentials, where the $S$-wave interaction has dominant contribution. For the $S$-wave $\Xi_c^{\prime} \bar D^*$ system, the operator $\mathcal{A}_{2}$ has connection to the interaction strength, i.e., $\langle\Xi_c^{\prime}\bar{D}^{*}|\mathcal{A}_{2}|\Xi_c^{\prime}\bar{D}^{*}\rangle=-2$ and $1$ with $J=1/2$ and $3/2$, respectively. Thus, we may conclude that the cutoff value of the $S$-wave isoscalar $\Xi_c^{\prime} \bar D^*$ state with $J^P=1/2^-$ should be smaller than that of the $S$-wave isoscalar $\Xi_c^{\prime} \bar D^*$ state with $J^P=3/2^-$ if getting the same binding energy. For the $S$-wave $\Xi_c \bar D^*$ system, there only exists the operator $\mathcal{A}_{1}$ with $\langle\Xi_c\bar{D}^{*}|\mathcal{A}_{1}|\Xi_c\bar{D}^{*}\rangle=1$ for the cases of $J=1/2$ and $3/2$. Therefore, the cutoff for the $S$-wave isoscalar $\Xi_{c}\bar D^*$ states with $J^P=1/2^-$ and $3/2^-$ has the same value for getting the same binding energy when performing the $S$-$D$ wave mixing analysis.}

{At present, there does not exist the experimental information involved in the coupling constant $g_{\sigma}$. Thus, we have to estimate the $g_{\sigma}$ value by the phenomenological model. 
Usually, there are two determined values for coupling constant $g_{\sigma}$, which are $0.76$ and $2.82$, by the spontaneously broken chiral symmetry \cite{Bardeen:2003kt} and the quark model \cite{Riska:2000gd}. Both of them were adopted in realistic study. As shown in Table \ref{gsu}, the loosely bound state solutions can be obtained with the cutoff values closed to 1 GeV for the $S$-wave isoscalar $\Xi_{c}^{\prime}\bar D^{*}$ state with $J^P=1/2^-$ by taking either $g_{\sigma}=0.76$ or $g_{\sigma}=2.82$. The uncertainty resulted from the $g_{\sigma}$ value can be compensated by smally changing $\Lambda$ value. Thus, the conclusion of existing the $S$-wave isoscalar $\Xi_{c}^{\prime}\bar D^{*}$ molecular state with $J^P=1/2^-$ is still hold when considering the uncertainties of the coupling constant $g_{\sigma}$.}

\renewcommand\tabcolsep{0.55cm}
\renewcommand{\arraystretch}{1.50}
\begin{table}[!htbp]
\centering
\caption{Bound state properties for the $S$-wave isoscalar $\Xi_{c}^{\prime}\bar D^{*}$ state with $J^P=1/2^-$ when considering the uncertainties of the coupling constant $g_{\sigma}$. Here, the cutoff $\Lambda$, binding energy $E$, and root-mean-square radius $r_{{\rm RMS}}$ are in units of $ \rm{GeV}$, $\rm {MeV}$, and $\rm {fm}$, respectively.}\label{gsu}
\begin{tabular}{cccc}\toprule[1.0pt]\toprule[1.0pt]
\multicolumn{4}{c}{$g_{\sigma}=0.76$}\\\midrule[1.0pt]
$\Lambda$  &$E$ &$r_{\rm RMS}$ &P(${}^2\mathbb{S}_{\frac{1}{2}}/{}^4\mathbb{D}_{\frac{1}{2}}$)\\\hline
0.92&$-0.50$ &4.02&\textbf{99.63}/0.37     \\
0.97&$-4.18$ &1.64&\textbf{99.45}/0.55     \\
1.02&$-11.90$ &1.06&\textbf{99.45}/0.55    \\\midrule[1.0pt]
\multicolumn{4}{c}{$g_{\sigma}=2.82$}\\\midrule[1.0pt]
$\Lambda$  &$E$ &$r_{\rm RMS}$ &P(${}^2\mathbb{S}_{\frac{1}{2}}/{}^4\mathbb{D}_{\frac{1}{2}}$)\\\hline
0.83&$-0.33$ &4.62&\textbf{99.71}/0.29\\
0.88&$-3.94$ &1.74&\textbf{99.53}/0.47\\
0.93&$-11.95$ &1.11&\textbf{99.51}/0.49\\
\bottomrule[1.0pt]\bottomrule[1.0pt]
\end{tabular}
\end{table}

{We should indicate that the $P_{cs}(4459)$ and $P_{\psi s}^\Lambda(4338)$ are also close to the corresponding thresholds of baryon-meson channels like the combination of the $\Lambda_c/\Sigma_c^{(*)}$ baryon and the $D_s^{(*)}$ meson.
Although the isovector $\Sigma_c^{(*)}\bar D_s^{(*)}$ channel cannot couple with the discussed isoscalar systems, the $\Lambda_c\bar D_s^{(*)}$ channel can interact with the discussed systems, which is not addressed in the present coupled channel analysis.}

\begin{figure}[htbp]
\centering
\includegraphics[width=8.7cm]{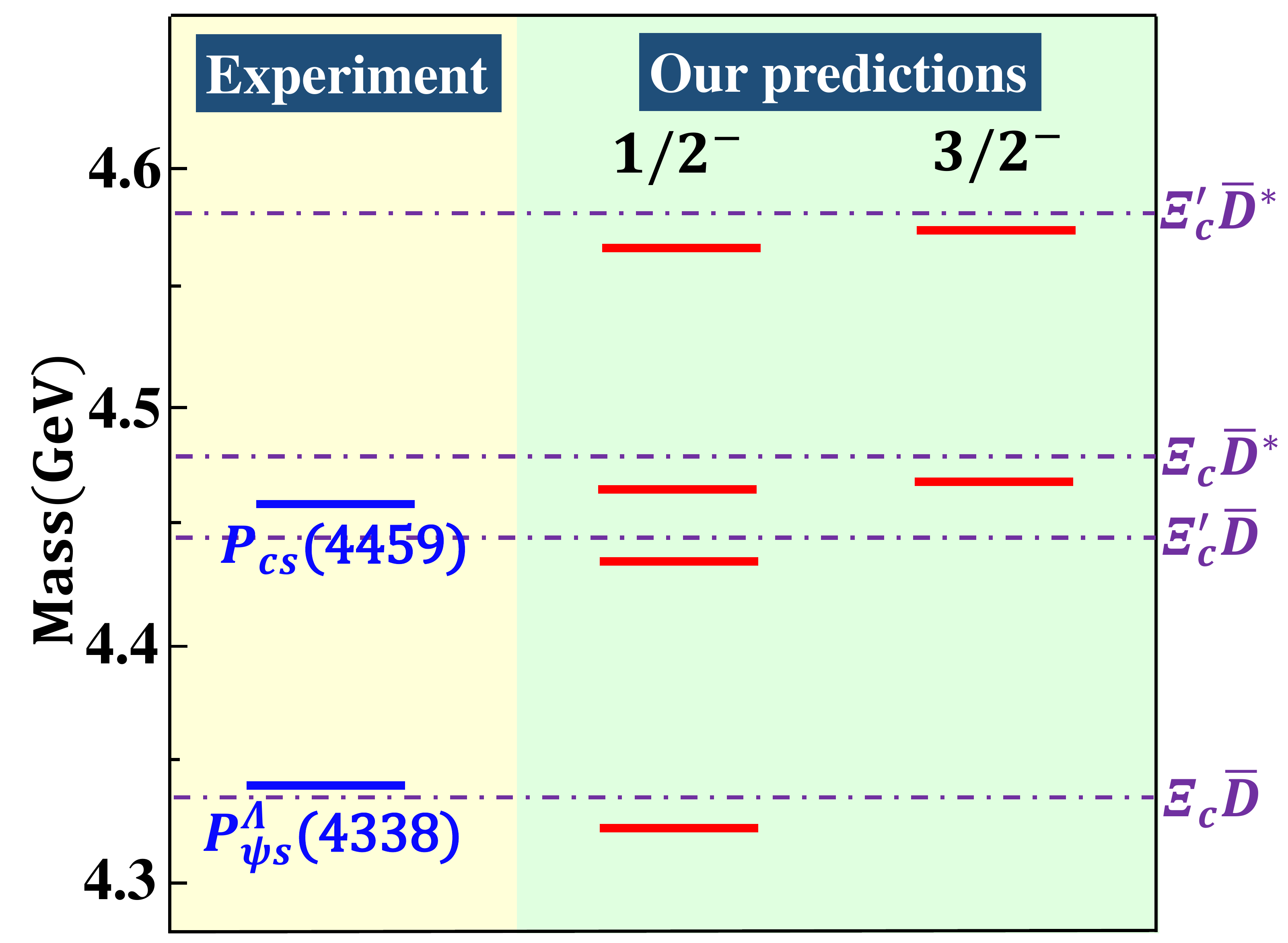}\\
\caption{The comparison of the masses between the observed hidden-charm pentaquarks with single strangeness \cite{Collaboration:2022boa,LHCb:2020jpq} and our predicted $S$-wave isoscalar $\Xi_c^{(\prime)}\bar D^{(*)}$-type hidden-charm molecular pentaquarks with single strangeness. Here, the purple dash-dotted lines denote the thresholds of these discussed hidden-charm molecular pentaquarks with single strangeness, and the blue and red thick solid lines represent the masses of the observed hidden-charm pentaquarks with strangeness and our predicted $S$-wave isoscalar $\Xi_c^{(\prime)}\bar D^{(*)}$-type hidden-charm molecular pentaquarks with single strangeness, respectively.}\label{EXPTHE}
\end{figure}

{By above discussion, we may conclude that there exists the characteristic mass spectra of the $S$-wave isoscalar $\Xi_c^{(\prime)}\bar D^{(*)}$-type hidden-charm molecular pentaquarks with single strangeness. In Fig. \ref{EXPTHE}, we compare the masses between the observed hidden-charm pentaquarks with single strangeness \cite{Collaboration:2022boa,LHCb:2020jpq} and our predicted $S$-wave isoscalar $\Xi_c^{(\prime)}\bar D^{(*)}$-type hidden-charm molecular pentaquarks with single strangeness.}

{In the following, we discuss the two-body hidden-charm decay behaviors of our predicted $S$-wave isoscalar $\Xi_c^{(\prime)}\bar D^{(*)}$-type hidden-charm molecular pentaquarks with single strangeness, which is inspired by the observation from LHCb of the $P_c$ and $P_{cs}$ states \cite{Aaij:2015tga,Aaij:2019vzc,Collaboration:2022boa,LHCb:2020jpq} via the two-body hidden-charm decay channels. In the past decades, the heavy quark symmetry was often used to discuss the properties of the hadrons which contain the heavy quarks. Thus, we may discuss the two-body hidden-charm decay behaviors of these predicted hidden-charm molecular pentaquarks with single strangeness within the heavy quark symmetry.}

\begin{figure}[!htbp]
\center
\includegraphics[width=7.5cm]{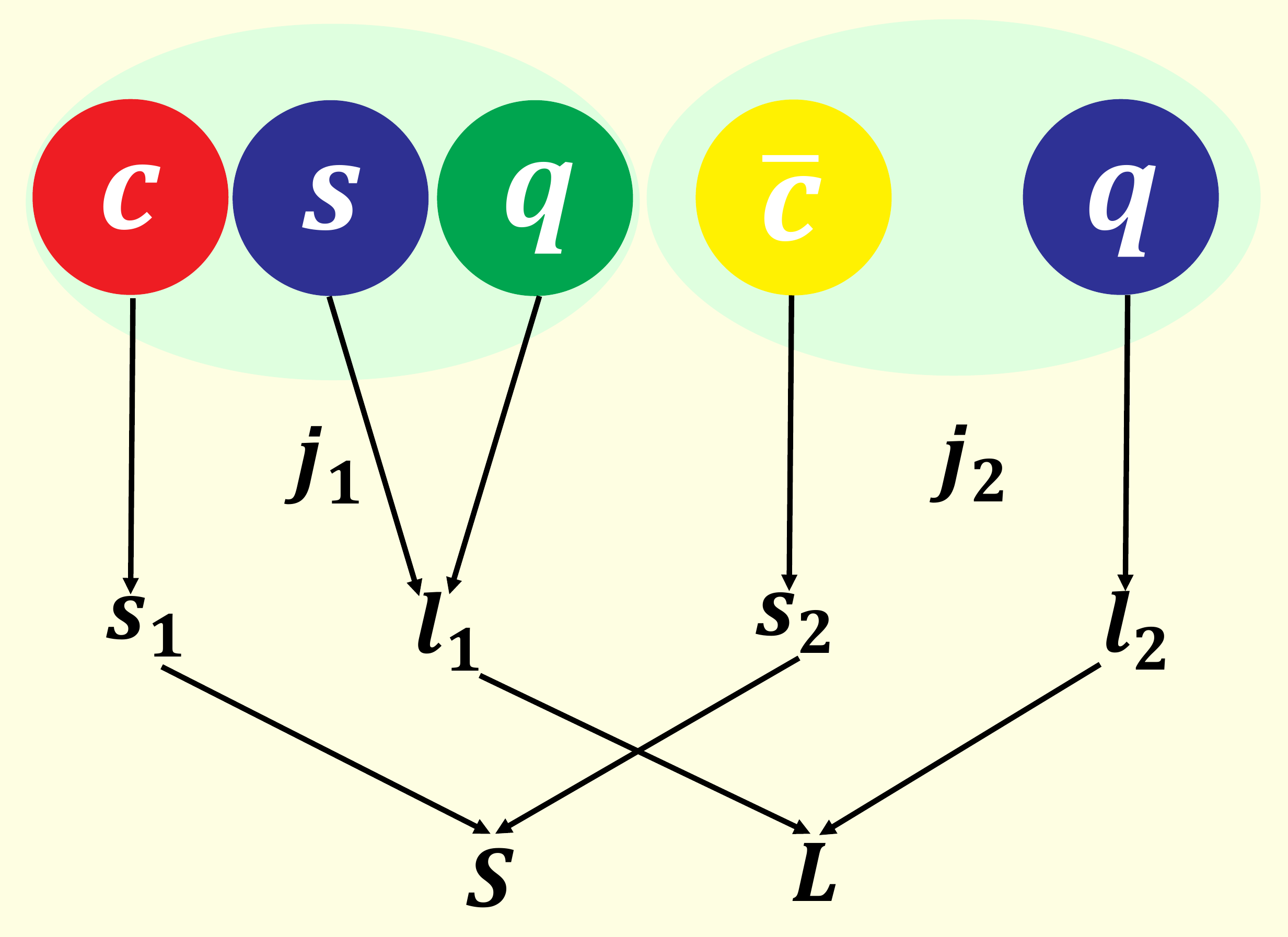}\\
\caption{Diagram for the calculation of the 9-$j$ coefficients. Here,  $j_1$ and $j_2$ denote the total angular momentum quantum numbers for the charmed baryons and anti-charmed mesons, respectively. $s_1$ and $s_2$ stand for the spin quantum numbers of the heavy quarks. $l_1$ and $l_2$ are the spin quantum numbers of the light quarks. And then, $l_1=0$ and $1$ for the $\Xi_c$ and $\Xi_c^{\prime}$ baryons, respectively.}\label{9j}
\end{figure}

{As shown in Fig. \ref{9j}, we expand the spin wave functions of the heavy hadron systems $|\ell_1 s_1 j_1, \ell_2 s_2 j_2, J M \big\rangle$ in terms of the heavy quark basis $|\ell_1 \ell_2 L, s_1 s_2 S, J M \big\rangle$, where the general relation can be expressed as
\begin{eqnarray}
\left|\ell_1 s_1 j_1, \ell_2 s_2 j_2, J M \right\rangle &=& \sum_{S,L}\hat{S} \hat{L} \hat{j_1} \hat{j_2}\left\{
\begin{array}{ccc}
\ell_1 & \ell_2 & L \\
s_1 & s_2 & S \\
j_1 & j_2 & J
\end{array}
\right\}\nonumber\\
&&\times|\ell_1 \ell_2 L, s_1 s_2 S, J M \big\rangle
\end{eqnarray}
with $\hat{A}=\sqrt{2A+1}$. Thus, we obtain
\begin{eqnarray}
|\Xi_{c} \bar D(J=\frac{1}{2})\big \rangle &=&\frac{1}{2}|S_{c\bar c}=0,\, L_{sqq}=\frac{1}{2},\, J=\frac{1}{2}\big \rangle\nonumber\\
&+&\frac{\sqrt{3}}{2}|S_{c\bar c}=1,\, L_{sqq}=\frac{1}{2},\, J=\frac{1}{2}\big \rangle,\\
|\Xi_{c} \bar D^*(J=\frac{1}{2})\big \rangle &=&\frac{\sqrt{3}}{2}|S_{c\bar c}=0,\, L_{sqq}=\frac{1}{2},\, J=\frac{1}{2}\big \rangle\nonumber\\
&-&\frac{1}{2}|S_{c\bar c}=1,\, L_{sqq}=\frac{1}{2},\, J=\frac{1}{2}\big \rangle,\\
|\Xi_{c}^{\prime} \bar D(J=\frac{1}{2})\big \rangle &=&\frac{1}{2}|S_{c\bar c}=0,\, L_{sqq}=\frac{1}{2},\, J=\frac{1}{2}\big \rangle\nonumber\\
&-&\frac{1}{2\sqrt{3}}|S_{c\bar c}=1,\, L_{sqq}=\frac{1}{2},\, J=\frac{1}{2}\big \rangle\nonumber\\
&+&\sqrt{\frac{2}{3}}|S_{c\bar c}=1,\, L_{sqq}=\frac{3}{2},\, J=\frac{1}{2}\big \rangle,\\
|\Xi_{c}^{\prime} \bar D^*(J=\frac{1}{2})\big \rangle &=&-\frac{1}{2\sqrt{3}}|S_{c\bar c}=0,\, L_{sqq}=\frac{1}{2},\, J=\frac{1}{2}\big \rangle\nonumber\\
&+&\frac{5}{6}|S_{c\bar c}=1,\, L_{sqq}=\frac{1}{2},\, J=\frac{1}{2}\big \rangle\nonumber\\
&+&\frac{\sqrt{2}}{3}|S_{c\bar c}=1,\, L_{sqq}=\frac{3}{2},\, J=\frac{1}{2}\big \rangle.
\end{eqnarray}
For these predicted  isoscalar $S$-wave $\Xi_c^{(\prime)}\bar D^{(*)}$-type hidden-charm molecular pentaquarks with single strangeness, the allowed two-body hidden-charm decay channels include the $J/\psi \Lambda$ and $\eta_c(1S) \Lambda$. In the heavy quark symmetry, we can estimate
\begin{eqnarray*}
\mathcal{R}_{\text{HQS}}^1&=&\frac{\Gamma[\Xi_c\bar D(1/2^-)\to\eta_c(1S)\Lambda]}{\Gamma[\Xi_c\bar D(1/2^-)\to J/\psi\Lambda]}=\frac{1}{3},\\
\mathcal{R}_{\text{HQS}}^2&=&\frac{\Gamma[\Xi_c\bar D^*(1/2^-)\to\eta_c(1S)\Lambda]}{\Gamma[\Xi_c\bar D^*(1/2^-)\to J/\psi\Lambda]}=3,\\
\mathcal{R}_{\text{HQS}}^3&=&\frac{\Gamma[\Xi_c^{\prime}\bar D(1/2^-)\to\eta_c(1S)\Lambda]}{\Gamma[\Xi_c^{\prime}\bar D(1/2^-)\to J/\psi\Lambda]}=3,\\
\mathcal{R}_{\text{HQS}}^4&=&\frac{\Gamma[\Xi_c^{\prime}\bar D^*(1/2^-)\to\eta_c(1S)\Lambda]}{\Gamma[\Xi_c^{\prime}\bar D^*(1/2^-)\to J/\psi\Lambda]}=\frac{3}{25}.
\end{eqnarray*}
For the $S$-wave $\Xi_{c}\bar D^*$ and $\Xi_{c}^{\prime}\bar D^*$ molecular states with $J^P={3}/{2}^{-}$, the $J/\psi \Lambda$ is the two-body hidden-charm decay channel via the $S$-wave coupling, while the $\eta_c (1S)\Lambda$ channel is suppressed since it is a $D$-wave interaction. By this effort, we can find several significant two-body hidden-charm decay channels for these predicted hidden-charm molecular pentaquarks with single strangeness, which may provide crucial information of searching for these predicted pentaquarks in future experiment. Besides the strong decay behavior of these predicted pentaquarks, their radiative decay behavior has aroused theorist's interest as given in Ref. \cite{Wang:2022tib}.}

\section{Discussion and conclusion}\label{sec4}

Very recently, the LHCb Collaboration announced the observation of the $P_{\psi s}^\Lambda(4338)$ in the $J/\psi\Lambda$ invariant mass spectrum of the $B^-\to J/\psi\Lambda\bar{p}$ process \cite{Collaboration:2022boa}. The $P_{\psi s}^\Lambda(4338)$ associated with the previously reported evidence of the $P_{cs}(4459)$ by LHCb shows a molecular-type characteristic mass spectrum of hidden-charm pentaquark with strangeness. Before that, a molecular-type characteristic mass spectrum of hidden-charm pentaquark was observed by discovering the $P_c(4312)$, $P_c(4440)$, and $P_c(4457)$ in the $J/\psi p$ invariant mass spectrum of the $\Lambda_b\to J/\psi p K$ \cite{Aaij:2019vzc}. In this work, we indicate the similarity of two molecular-type characteristic mass spectra mentioned above.

\begin{figure}[htbp]
\centering
\includegraphics[width=8.7cm]{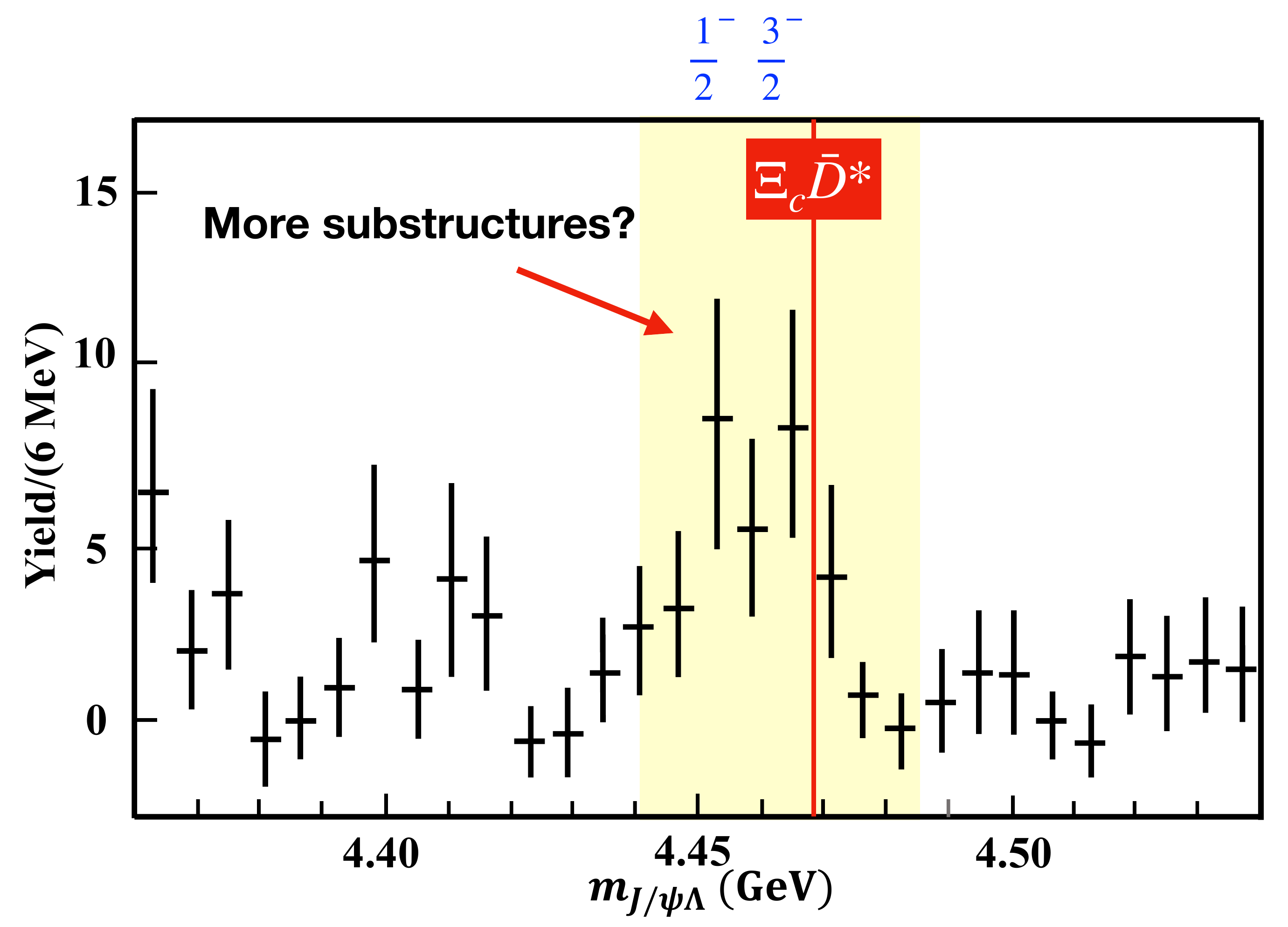}\\
\caption{The possible evidence  of exiting substructures contained in the $P_{cs}(4459)$ enhancement structure. Here, the data of $J/\psi\Lambda$ invariant mass spectrum are from the LHCb \cite{LHCb:2020jpq}. The red solid line is the $\Xi_c\bar{D}^*$ threshold. Near this threshold, there should exist two $\Xi_c\bar{D}^*$ molecular states with $J^P=1/2^-$ and $3/2^-$. The event cluster in the yellow ban shows possible double-peak evidence but it is not obvious, which can be tested by more precise data. }
\label{fig4}
\end{figure}

For quantitatively depicting this characteristic mass spectrum of hidden-charm pentaquark with strangeness, we apply the OBE model to obtain the interactions of the $\Xi_c\bar{D}^{(*)}$ systems, where the effective potentials of the focused systems can be deduced. By solving the Schr\"odinger equation, we find out the bound state solutions of the $\Xi_c\bar{D}^{(*)}$ pentaquark systems. The numerical result shows the existence of two $\Xi_c\bar{D}^*$ molecular pentaquarks with $J^P=1/2^-$ and $3/2^-$, which can be related to the $P_{cs}(4459)$ structure. In fact, there exists a similar situation to the $P_c(4450)$, which was observed firstly by LHCb in 2015 \cite{Aaij:2015tga}. When reanalyzing more precise data of the $\Lambda_b\to J/\psi p K$ decay in 2019, LHCb indicated that the $P_c(4450)$ structure contains two substructures $P_c(4440)$ and $P_c(4457)$ \cite{Aaij:2019vzc}. We may conjecture that the double-peak phenomenon can be happened to the $P_{cs}(4459)$ structure\footnote{In former experimental analysis \cite{LHCb:2020jpq}, LHCb tried to fit the enhancement structure corresponding to the $P_{cs}(4459)$ by two substructures \cite{LHCb:2020jpq}. Due to the limit of precision of data, LHCb cannot make firm conclusion of the existence of substructure phenomenon \cite{LHCb:2020jpq}.} as illustrated in Fig. \ref{fig4}. Thus, we strongly suggest the experimental colleagues to reanalyze the $J/\psi\Lambda$ invariant mass spectrum of $\Xi_b^-\to J/\psi \Lambda K$ based on more precise data collected with the running of high-luminosity LHC.

{If checking the resonant parameter of the reported $P_{\psi s}^{\Lambda}(4338)$ \cite{Collaboration:2022boa}, we cannot ignore a fact that the central value of the $P_{\psi s}^{\Lambda}(4338)$ mass is a little bit larger than the threshold of the $\Xi_c\bar{D}$ channel. Thus, it seems unnatural to assign the $P_{\psi s}^{\Lambda}(4338)$ as the $\Xi_c\bar{D}$ molecular state with $J^P=1/2^-$ since the $\Xi_c\bar{D}$ molecular state has negative binding energy. As shown in Fig. \ref{fig1}, the mass gap between the $P_{cs}(4459)$ and the $P_{\psi s}^\Lambda(4338)$ is smaller than that between the $P_c$ states. If considering the similarity of the mass gap between the $P_c$ states and that between the $P_{cs}$ states. Taking the mass gap $137$ MeV involved in the $P_c$ states and the mass of the $P_{cs}(4459)$ as input, the $P_{cs}$ partner of the $P_c(4312)$ should be located at $4322$ MeV, which is lower than the observed $P_{\psi s}^\Lambda(4338)$ by LHCb via the $B^-\to J/\psi \Lambda \bar{p}$ process. Thus, with more precise data, we also suggest to check such scenario in future experiment like LHCb. Here, a high-priority task is to confirm the observation of the $P_{\psi s}^{\Lambda}(4338)$ by the $\Xi_b^-\to J/\psi \Lambda K$ process.}

As a prediction, in this work we also give another molecular-type characteristic mass spectrum of hidden-charm $\Xi_c^{\prime}\bar{D}^{(*)}$ pentaquark systems. Our result shows that there exist a $\Xi_c^{\prime}\bar{D}$ molecular pentaquark with $J^P=1/2^-$ and two $\Xi_c^{\prime}\bar{D}^*$ molecular pentaquarks with $J^P=1/2^-$ and $3/2^-$, which are near the thresholds of $\Xi_c^{\prime}\bar{D}$ and $\Xi_c^{\prime}\bar{D}^{*}$, respectively. Thus, searching for this predicted molecular-type characteristic mass spectrum of the $\Xi_c^{\prime}\bar{D}^{(*)}$ pentaquark systems will be a new task for further experimental exploration of pentaquarks.

\vfil

\noindent{\it Note added.--}When preparing the present paper, we noticed a similar work  from Karliner and Rosner \cite{Karliner:2022erb} appeared in arXiv. They also indicated the similarity of the mass gaps for these reported $P_c$ and $P_{cs}$ states.
Different from Ref. \cite{Karliner:2022erb}, in this work we addressed a dynamical calculation within the OBE model to illustrate the importance of the molecular-type characteristic mass spectra of hidden-charm $\Xi_c^{(\prime)}\bar{D}^{(*)}$ pentaquark systems.

\section*{Acknowledgement}
This work is supported by the China National Funds for Distinguished Young Scientists under Grant No. 11825503, National Key Research and Development Program of China under Contract No. 2020YFA0406400, the 111 Project under Grant No. B20063, and the National Natural Science Foundation of China under Grant No. 12175091.

\end{document}